\documentclass[fleqn,usenatbib]{mnras}

\usepackage{newtxtext,newtxmath}

\usepackage[T1]{fontenc}

\DeclareRobustCommand{\VAN}[3]{#2}
\let\VANthebibliography\thebibliography
\def\thebibliography{\DeclareRobustCommand{\VAN}[3]{##3}\VANthebibliography}

\usepackage{graphicx}	

\title[Distant emission clouds in merging galaxies]{The TELPERION survey for extended emission regions around AGN: a 
strongly-interacting and merging galaxy sample}

\author[W.C. Keel et al.]{
William C. Keel$^{1,2}$ \thanks{E-mail: wkeel@ua.edu},
Alexei Moiseev$^{3,4}$,
Roman Uklein$^{3}$,
and Aleksandrina Smirnova$^{3}$
\\
$^{1}$Dept. of Physics and Astronomy, University of Alabama, Box 870324, Tuscaloosa, AL 35487, USA\\
$^{2}$SARA Observatory, Embry-Riddle Aeronautical University, Daytona Beach FL 32114, USA \\
$^{3}$Special Astrophysical Observatory, Nizhny Arkhyz, Russia\\
$^{4}$Lomonosov Moscow State University, Sternberg Astronomical Institute, Universitetsky pr. 13, Moscow 119234, Russia\\
}

\date{Accepted XXX. Received YYY; in original form ZZZ}

\pubyear{2023}

\begin{document}
\label{firstpage}
\pagerange{\pageref{firstpage}--\pageref{lastpage}}
\maketitle

\begin{abstract}
We present the results of a search for Extended Emission-Line Regions (EELRs) ionized by extant or
recently-faded active galactic nuclei (AGN), using [O III] narrowband imaging and
spectroscopic followup. The sample includes 198 galaxies in 92 strongly interacting or
merging galaxy systems in the range $z=0.009-0.0285$. Among these, three have
EELRs extended beyond 10 kpc in projection from the nucleus detected in previous studies. We
identify a single new distant emission region, projected 35 kpc from UGC 5941. Our optical spectrum does not 
detect He II, but its strong-line ratios put this in the 
same class as securely-characterized EELR clouds. The nucleus of UGC 5941 is dominated by recent star formation, preventing
detection of any weak ongoing AGN.
Overall counts of distant EELRs in this and the previous TELPERION samples give
incidence 2--5\% depending on galaxy and AGN selection, 20-50 times higher
than the Galaxy Zoo EELR survey with its higher surface-brightness threshold
and much larger input sample. AGN in interacting and merging
systems have an increased detection rate $12\pm 6$\%, while none are detected around noninteracting AGN. Some of these
AGN are at luminosity low enough to require additional X-ray or far-infrared information to tell whether the EELR ionization level
suggests long-term fading.
\end{abstract}




\section{Introduction}
Among active galactic nuclei (AGN), most types are characterised by strong emission lines, whose ionization level distinguishes their spectra
from those of objects powered by star formation. In addition to the high-density broad-line region and surrounding narrow-line
region, some AGN also have extended emission-line regions (EELRs) extending for kiloparsecs or tens of kiloparsecs from the nucleus itself.

Objects described as EELRs in the literature are ionized directly or indirectly by AGN, as distinct from other kinds of extended emission-line regions
such as starburst winds, widespread star formation, or shocks occurring in merging galaxies. EELRs may be photoionized by radiation from the AGN, or
ionized by the effects of jets or broader outflows from radio-loud AGN, generally including shocks and associated with emission-line Doppler widths $>100$ km s$^{-1}$
\citep{StocktonCanalizo}.

Photoionized EELRs offer the opportunity to learn about the AGN's ionizing radiation in directions or at times we cannot sample
directly. They often occur in pairs, triangular in projection and pointing to the AGN (ionization cones, reviewed by
\citealt{Wilson}), which furnish strong evidence in favor of a broad picture of circumnuclear obscuration producing much of the difference between type 1 and 2 AGN 
(depending on whether the broad-line region is obscured or not along our line of sight). A second facet of photoionized EELRs was exemplified by the discovery of Hanny's Voorwerp \citep{Voorwerp} during the initial Galaxy Zoo project \citep{Lintott2008}, where the ionizing luminosity
needed to match the EELR's luminosity and ionization level far exceed what is seen from the host galaxy's nucleus directly. 
This provides the possibility of sampling the AGN luminosity at times up to $10^5$ years earlier than the
direct view, giving the promise of retrieving the accretion history of supermassive black holes over timescales not otherwise 
possible on human timescales. The characterization of Hanny's Voorwerp led to a followup search by Galaxy Zoo
volunteers, based on the distinctive colour of low-redshift [O III] emission regions in $gri$ colour-composite renditions of
Sloan Digital Sky Survey (SDSS) images. This project confirmed another 19 distant EELRs, most found for the first time, and allowed a
rough estimate (0.2--2$\times 10^5$ years) of the duration of luminous accretion events for these AGN \citep{Keel2012}.
About 0.12\% of the AGN examined were found to show distant EELRs to surface-brightness levels apparent in SDSS data.

To complement the Galaxy Zoo searches, we have carried out a phased survey called
TELPERION\footnote{An acronym for Tracing Emission Lines Probing Extended Regions Ionized by Once-active Nuclei, and recalling a past age of vanished brilliance for readers of J.R.R. Tolkien's stories of the First Age of Middle-Earth.}.
This uses narrowband imaging of several hundred galaxies in a convenient redshift range to seek emission clouds at lower surface brightness than
was detectable in the original SDSS images, out to projected distances in principle as much as 250 kpc from the AGN hosts (so,
for photoionized EELRs, we
could {detect clouds tracing the luminosity of the AGN as much as $10^5$ years before the directly-observed epoch).
TELPERION includes samples of galaxies which are complete in various senses. The initial phase of the survey \citep{Knese}
was comprised of Seyfert galaxies with known H I properties, so it was clear which systems had enough neutral gas to
produce detectable [O III] emission. This phase uncovered one distant EELR, Mkn 1, out of 26 AGN hosts. A second survey
phase \citep{TELPERION} targeted 111 luminosity-selected AGN hosts (and any companions), finding distant EELRs
in NGC 235 and NGC 5514. This phase also included merging galaxies from an enhanced version of the ``Toomre sequence" 
of ongoing mergers in our redshift range, recovering the EELR previously found in NGC 7252 by \cite{Schweizer}.

Our primary interest in these surveys is identification of EELRs extending beyond 10 kpc  in projection from the AGN,
which we term distant EELRs. Compared to regions smaller and closer to the galaxy nuclei, these are less likely to consist of the ordinary ISM of the host 
galaxy, allow us to probe
longer light-travel-time delays, and are easier to detect outside the bright parts of host galaxies for
better completeness of our samples. We also note that the great majority of distant EELRs in both the Galaxy Zoo sample and TELPERION
are found around radio-quiet AGN (simply because most AGN are not radio-loud objects with kpc-scale jets). Radio-loud objects
have EELRs that often reflect outflow associated with the jets or shocks as the jets encounter the host galaxy ISM, rather than
the more kinematically-quiescent AGN-photoionized clouds we focus on.

A recurring pattern among the distant EELRs found in both the Galaxy Zoo and TELPERION surveys has been that
almost all are found in interacting or merging galaxies, possibly because tidal features often include neutral gas
that can be observed when ionized by escaping UV continuum from the AGN (and such features often depart from the
host galaxy's plane, reducing the absorption along the path of the ionizing photons). We present here results of
a further phase of the TELPERION survey specifically aimed at interacting and merging galaxies, and preliminary 
statistics of the occurrence of distant EELRs among all four sample phases of the TELPERION survey.

Distance-dependent quantities are calculated using ${\rm H_0}=69.6$ km s$^{-1}$ Mpc$^{-1}$, $\Omega_{\rm M}=0.286$,
and assuming flat geometry \citep{H0}, with the Javascript calculator described by \cite{Wright}.

\section{Sample Selection and Observations}

\subsection{Sample selection}
\label{sec:sample} 

To further examine the connection between distant EELRs and strong interactions or mergers, we selected galaxies
from the \cite{Darg} sample, of galaxies having been flagged as mergers in the Galaxy Zoo project \citep{Lintott2008},
which classified galaxies from data release 7 (DR7) of the SDSS \citep{DR7}.
This sample includes ongoing mergers, tidally distorted interacting systems, and a small admixture of projected pairs at
substantially different distances \citep{overlapcatalog}.
As in the previous TELPERION phases, we used off-the-shelf narrowband filters to detect [O III] $\lambda 5007$-\AA\
emission (hereinafter simply [O III])  within the range $z=0.0085-0.0285$, a range which includes 217 systems from the \cite{Darg} catalog.
We did not require the galaxy nucleus to be classified as an AGN from optical spectra, but we did omit galaxies with central starbursts as judged from
SDSS spectra, to avoid emission from outlying star-forming regions or starburst winds, not associated with an AGN. Following \cite{StasinskaLeitherer},
\cite{CidFernandes}, and \cite{Bergvall}, we omitted galaxy pairs where a non-AGN nuclear spectrum has H$\beta$ equivalent
width EW(H$\beta) > 20$ \AA. This criterion helps make our EELR survey more efficient, though we cannot apply it completely,
since some pairs have only a single SDSS spectrum available, and some galaxies' spectra represent off-nuclear knots rather than
their nuclei.
This left 95 systems, most with two members.
Of these, three systems are found to be projections of unrelated and undistorted galaxies based on redshifts in later SDSS data releases, 
so we omit these from our analysis:
MCG +06-20-035, SDSS J114328.50+094947.2, and SDS J140531.65-012147.9. IC 3935 includes a
background system, but the foreground disc is asymmetric across a wide radial range so we include
it as the aftermath of a strong interaction or minor merger. Our sample then includes 92 strongly-interacting and merging systems
containing 198 distinct galaxies (although what counts as ``distinct" in an ongoing merger is sometimes
a matter of individual judgment).
 
The \cite{Darg} sample of interacting and merging galaxies was selected using classifications of galaxies in SDSS DR7. One system observed in our
sample,}
NGC 5541 was included as a photometric object (SDSS J141631.80+393520.5) in SDSS DR7 but not DR9 \citep{DR9} and later, apparently because of sky-subtraction issues due to the bright star HD 125111 projected 9\arcmin \ away. A coordinate-based match cross-match of sample objects between DR7 and DR9 would miss this one.

To help understand the content of the sample further, we classified the nuclear spectra of the galaxies based on SDSS spectra and our previous observations 
\citep{KKHH}, for simplicity using the [O III]/H$\beta$-[N II]/H$\alpha$  BPT diagram (\citealt{BPT}, \citealt{Kewley2001}, \citealt{Kauffmann}) for emission-line spectra. The pipeline classifications for AGN listed in the SDSS
database use a linear approximation to the Seyfert+LINER lower bound which falls above the maximum-starburst curves from \cite{Kewley2001}, and \cite{Kauffmann}
everywhere, so we work from the SDSS line fluxes when model-corrected for underlying stellar absorption (and apply color-based corrections for H$\alpha$ and
H$\beta$ absorption for our own data). From emission lines, we find 68 star-forming objects,
32 LINERS, 11 Seyfert galaxies including objects near the Seyfert/LINER boundary, and 3 composite nuclei near the AGN lower boundary.
Four nuclei show absorption spectra with strong Balmer lines, indicating post-starburst systems. We find 37 quiescent systems, with the equivalent width of H$\alpha$ emission
less than 3 \AA\  and no other emission detected except in some cases [O II]. No optical spectra were available for 43 of the 198 galaxy nuclei
in this sample.

\subsection{Imaging observations}

Our imaging procedures followed those in the previous phases of the TELPERION
survey (\citealt{Knese}, \citealt{TELPERION}). We searched for extended and distant
[O III] features using a filter with center wavelength 510 nm and nominal half-transmission points at 505 and 515 nm
(denoted F510), which should shift blueward in a converging beam by no more than 4 \AA\ in our telescopes. 
Our sample includes galaxies where the stronger [O III] emission line is observed within
this range ($z=0.0085-0.0285$). Our narrowband images were obtained with the remotely-operated 1-m telescopes of the SARA
observatory \citep{SARA} at Kitt Peak (KP, 0.44\arcsec per pixel, field of view 14.5\arcmin \ square) and La Palma (the Jacobus Kapteyn Telescope, JKT;
0.34\arcsec per pixel, field of view 11.6\arcmin \ square). While
explicit continuum subtraction often has large residuals in star images due to PSF mismatch, we did 
obtain $V$ images for comparison, with one of those telescopes or the 0.6m SARA instrument at Cerro Tololo
(CT). Our survey F510 images typically reach a surface brightness limit $1.0 \times 10^{-16}$ erg cm$^{-2}$ s$^{-1}$ arcsec$^{-2}$ 
in the $\lambda 5007$ line. The $V$ continuum images were exposed to have at least twice the signal-to-noise ratio of
the narrowband data. As noted with the previous TELPERION samples, use of [O III] for such a survey helps suppress detections
not only of star-forming regions relative to AGN-ionized gas, but other emission clouds not ionized by AGN,
including large-scale shocks in merging systems and low-ionization material
in the cores of galaxy clusters, in which the strongest optical emission lines are typically H$\alpha$ and [N II].

The imaging observations are documented in Table \ref{tbl-obslog}, with observing site and UTC date for each observation. 
When poor image quality led us to repeat an
observation, only the higher-quality one is listed. Images were examined with various levels of Gaussian and median smoothing in 
search of diffuse [O III] structures. Potential diffuse [O III] clouds were
reobserved for confirmation before performing spectroscopic followup, to reduce time spent on
false positives near the detection threshold (effectively allowing the survey to reach a deeper limiting surface brightness).

For the previously unknown distant [O III] cloud near UGC 5941, we obtained a
total of 13 one-hour exposures in F510 to examine its structure, at both KP and the JKT.

Images were acquired for this project over the time frame from January 2017 to May 2022. 
It was interrupted not only by shutdowns associated with unavailability of onsite support personnel
during the CoViD-19 pandemic lockdowns at each site, but the Cumbre Vieja volcanic eruption on the island of La Palma
(September 2021 - January 2022) and the aftermath of the Contreras wildfire at
Kitt Peak (June 2022 - January 2023).

\begin{figure*}
	\includegraphics[width=140mm,angle=90]{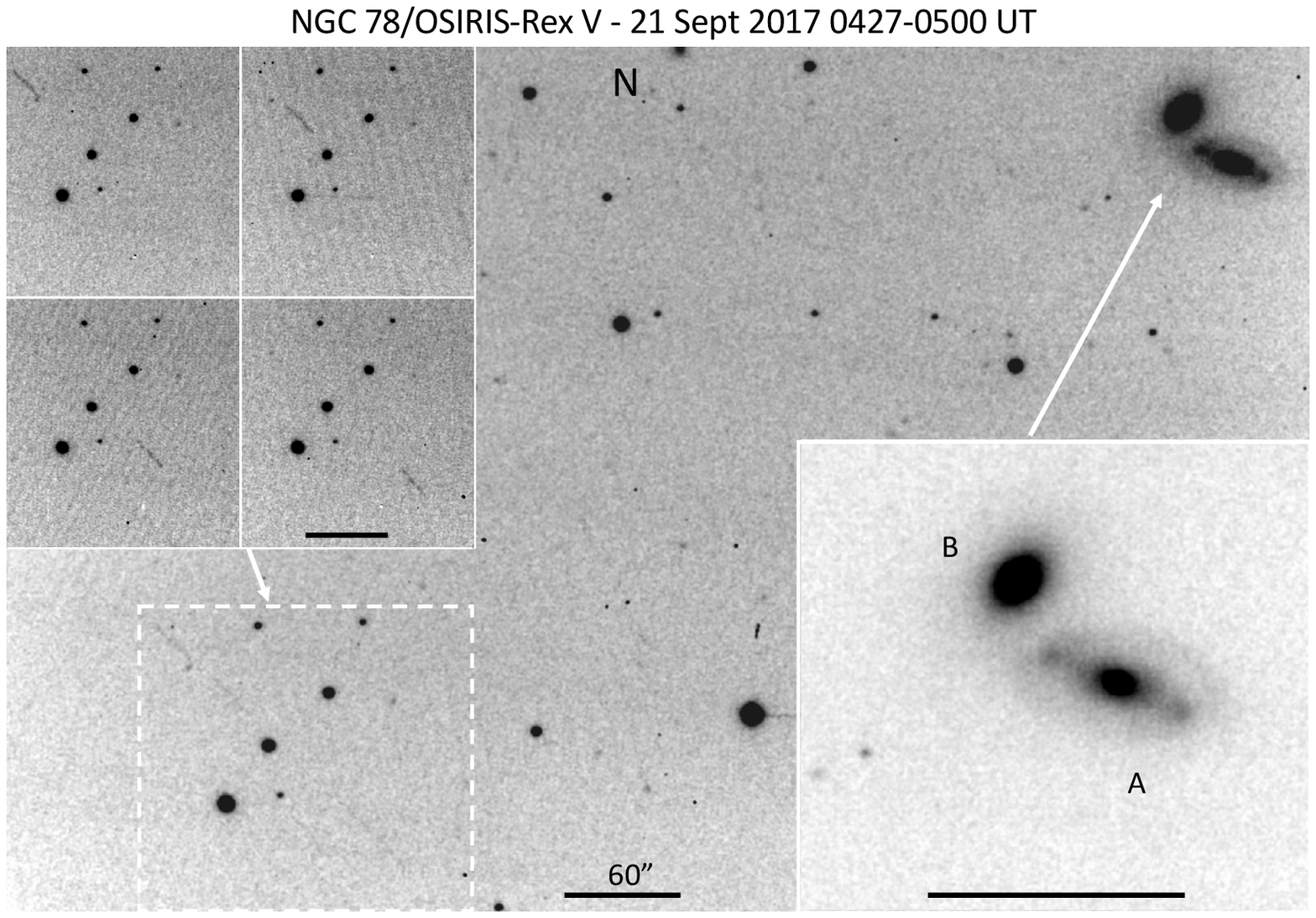}
    \caption{$V$-band continuum image of the galaxy pair NGC 78AB along with the OSIRIS-Rex asteroid probe during Earth swingby,
    obtained with the Jacobus Kapteyn Telescope. The
    main panel shows a combination of four 5-minute exposures, combined with the unusual procedure of taking the maximum value at each pixel location
    to retain the spacecraft trails. The four insets show the trails during each individual exposure, and the right-hand inset shows the galaxy
    pair at lower contrast to emphasize inner detail. The scale bar in each case is 60\arcsec in length. 
    The range to OSIRIS-Rex was 837,300 km in the middle of this sequence.
    }
    \label{fig:NGC78-ORex}
\end{figure*}

These images were not as confused by asteroid trails as the previous TELPERION phase. However, attention to the order of targets
during one night allowed us to include the OSIRIS-REx asteroid probe in the $V$ image of the NGC 78 system during an Earth swingby. 
We illustrate this, along with the $V$ continuum image of this galaxy pair, in Fig. \ref{fig:NGC78-ORex}. Coordinates and spacecraft
range were calculated using the JPL Horizons ephemeris service\footnote{https://ssd.jpl.nasa.gov/horizons/app.html\#/} .  

\subsection{Spectroscopic followup}
The single candidate EELR newly found in this survey, near UGC 5941, was observed spectroscopically using the SCORPIO-2 multimode 
instrument \citep{SCORPIO-2} on the 
6-meter telescope (Bol'shoi Teleskop Azimutal'nyi, BTA) at the Special Astrophysical Observatory of the Russian Academy of Sciences.
The SCORPIO-2 instrument was operated in its long-slit spectroscopic mode, employing a 1\arcsec \ slit with a usable length of 394\arcsec, 
oriented in position angle 39$^\circ$ through the nucleus of UGC 5941.
We obtained a 4500s exposure on 2023 March 15 UTC with the E2V 261-84 CCD detector \citep{Afanasieva}. 
The spectrum covered the range
3930--8560 \AA\ with 1.21 \AA\  pixel sampling (reciprocal dispersion). As well as the potential EELR, the spectrum records fainter emission
from a region just to the SW of the galaxy disc. Since the [S II] lines overlap the spectral range affected by telluric B-band absorption,
a correction was applied based on the spectrum of a reference star. The sensitivity of this spectrum detects a larger extent to
the cloud than even our deepest stacked direct image, showing [O III] for 9\arcsec along the slit and two emission peaks.

We derive error and upper-limit estimates by adding fictitious lines at random wavelengths while simultaneously fitting multiple Gaussian profiles in each wavelength range (using the IRAF {\tt splot} task), so these estimates include the effects of line width as derived from the strong lines and
nonlinear effects of the fitting algorithm.

We also incorporate long-slit data from SCORPIO-2 in three slit positions covering the emission filaments in NGC 5278/9, as
described in \cite{crossion}.


\subsection{Additional narrowband images}\label{section:mangal}

We make use of further narrowband [O III] images of two previously-detected EELR systems, Mkn 266 and NGC 5278/9. On- and off-band images
of NGC 5278/9 for [O III] were obtained with the MaNGaL tunable-filter system \citep{MaNGaL} at the 2.5 telescope of the Caucasus
Mountain Observatory\footnote{http://lnfm1.sai.msu.ru/kgo/main.php} (\citealt{KGO}, \citealt{KGOoptics}), with 2400s exposures for each.
A scaled difference image was used to isolate pure [O III] emission. For Mkn 266, we
examine an [O III] image produced by line fits through a data cube obtained with SCORPIO-2 in its Fabry-Perot scanning mode
\citep{FPIreview}, with total exposure 3600s.

\begin{table*}
	\centering
 \begin{minipage}{140mm}
	\caption{Sample galaxies and log of imaging observations}
	\label{tbl-obslog}
	\begin{tabular}{lcccccll} 
		\hline
SDSS name	& $z$ & 	V date & V site	&	F510 date	& F510 site	&	Other ID	&	Notes	\\ 
		\hline
SDSS J001755.39+002006.8 	& 0.018 & 	2017-11-16 & 	CT	&	2022-01-20 & 	KP	&	MCG +00-01-059	&	\\ 
SDSS J002027.50+005000.8 	& 0.018 & 	2017-09-20 & 	JKT	&	2021-10-06 &	KP	&	NGC 78	&		\\ 
SDSS J014503.30-001805.4 	& 0.018 & 	2017-01-17 &	JKT	&	2020-12-21 & 	KP	&	UGC 1225	&		\\
SDSS J020920.84-100759.1 	& 0.013 & 	2017-09-20 & 	CT	&	2017-12-20 & 	KP	&	NGC 833	&		\\
SDSS J024213.22-082934.9 	& 0.024 & 	2017-09-28 & 	JKT	&	2020-01-01 & 	KP	&	KUG 0239-087	&		\\ 
SDSS J032045.32-010314.2 	& 0.021 & 	2017-10-31 & 	CT	&	2018-01-17 & 	KP	&	UGC 2691	&	\\ 
SDSS J075121.01+501409.8 	& 0.024 & 	2017-11-13 & 	KP	&	2021-12-08 & 	KP	&	UGC 4052	&		\\ 
SDSS J080724.81+065147.5	& 0.015 & 	2017-11-20 & 	CT	&	2021-02-10 & 	KP	&		&		\\ 
SDSS J080945.62+255253.2 	& 0.025 &	2018-03-13 & 	JKT	&	2018-03-12 & 	KP	&	IC 2229	&		\\ 
SDSS J081312.31+402510.3	& 0.026 &	2018-03-15 & 	JKT	&	2018-12-13 &	KP	&		&		\\ 
SDSS J082724.89+233402.1	& 0.019 & 	2018-03-15 & 	JKT	&	2019-01-14 & 	KP	&		&		\\ 
SDSS J083031.67+181221.9	& 0.027 &	2017-11-20 & 	CT	&	2022-03-31 & 	KP	&		&		\\ 
SDSS J083824.01+254516.2 	& 0.018 &	2019-03-06 & 	JKT	&	2019-03-04 & 	KP	&	NGC 2623	&		\\ 
SDSS J084740.18+255336.6	& 0.022 & 	2018-01-17 & 	KP	&	2022-04-05 &	KP	&	UGC 4597	&		\\ 
SDSS J084924.19+190427.4 	& 0.013 & 	2018-05-15 & 	JKT	&	2020-12-21 & 	KP	&	NGC 2673	&		\\ 
SDSS J085601.86+421624.6 	& 0.028 & 	2018-11-15 & 	JKT	&	2020-02-12 & 	KP	&	MCG +07-19-002	&		\\
SDSS J090159.93+352031.9	& 0.026 & 	2019-12-17 & 	JKT	&	2020-02-12 & 	KP	&		&		\\ 
SDSS J090554.49+471045.5	& 0.027 &	2021-03-17 & 	JKT	&	2021-03-17 & 	JKT	&	 UGC 4765	&		\\ 
SDSS J091029.79+203344.3 	& 0.028 &	2019-12-27 & 	JKT	&	2020-12-14 &	JKT	&		&		\\ 
SDSS J091646.06+454849.6 	& 0.026 &	2019-02-04 & 	JKT	&	2020-12-21 &	KP	&	MCG +08-17-066	& \\
SDSS J092229.43+502548.4 	& 0.027 &	2019-11-25 &	JKT	&	2021-02-10 & 	KP	&	MCG +08-17-087	&		\\ 
SDSS J092540.12+221920.8 	& 0.025 &	2019-11-26 &	JKT	&	2021-02-16 & 	KP	&	CGCG 121-087 	&		\\ 
SDSS J093456.42+384441.9	& 0.015 &	2019-11-26 & 	JKT	&	2021-03-10 &	KP	&		&		\\
SDSS J093634.03+232627.0 	& 0.028 &	2021-03-17 &	JKT	&	2021-03-17 & 	JKT	&	CGCG 122-029	&		\\
SDSS J093939.93-001407.0 	& 0.016 &	2021-03-12 & 	JKT	&	2020-02-24 & 	KP	&	NGC 2951	&		\\ 
SDSS J094751.71+084950.8 	& 0.018 & 	2019-11-26 &  	JKT	&	2021-03-10 &	KP	&	MCG +02-25-036	&		\\
SDSS J095648.41+530853.8 	& 0.025 &	2018-03-13 &  	JKT	&	2019-04-24 &	KP	&	VV 740	&		\\
SDSS J100358.83+221633.8 	& 0.020 &	2018-05-15 & 	JKT	&	2020-12-21 & 	KP	&	UGC 5420	&		\\ 
SDSS J100549.83+003800.0  	& 0.021 &	2019-12-27 & 	JKT	&	2020-01-06 & 	KP	&	IC 590	&		\\
SDSS J102424.36+280143.3  	& 0.021 &	2019-12-27 & 	JKT	&	2020-02-12 &	KP	&	NGC 3232	&		\\
SDSS J102743.61+011138.2	& 0.022 &	2019-12-27 & 	JKT	&	2021-02-16 & 	KP	&		&		\\ 
SDSS J103635.16+374228.6	& 0.023 &	2019-12-27 & 	JKT	&	2021-03-17 &	JKT	&		&		\\ 
SDSS J104503.59+002603.6	& 0.026 &	2019-12-27 & 	JKT	&	2020-12-14 & 	JKT	&		&		\\ 
SDSS J104646.71+030942.8	& 0.022 &	2019-12-27 & 	JKT	&	2022-04-04 & 	KP	&		&		\\ 
SDSS J105021.59+412750.5 	& 0.024 &	2018-03-13 & 	JKT	&	2019-05-04 &  	KP	&	UGC 5941	& New [O III] cloud	\\ 
SDSS J105143.63+510119.7 	& 0.025 &	2019-12-27 & 	JKT	&	2020-02-24 &  	KP	&	NGC 3406	&	\\ 
SDSS J105237.39+080024.1 	& 0.022 &	2021-03-12 & 	JKT	&	2022-01-10 & 	KP	&	ALFALFA 5-358	&		\\ 
SDSS J105738.52+410317.6	& 0.025 &	2020-01-14 & 	JKT	&	2021-12-08 & 	KP	&		&		\\  
SDSS J111809.78+302424.3 	& 0.026 &	2021-03-12 &	JKT	&	2021-02-10 &  	KP	&	UGC 6314	&		\\ 
SDSS J113218.80+555540.8 	& 0.021 &	2020-01-14 &  	JKT	&	2021-12-09 &  	KP	&		&		\\ 
SDSS J113240.25+525701.4 	& 0.026 &	2018-05-15 & 	JKT	&	2018-12-13 & 	KP	&	Mkn 176	&		\\ 
SDSS J114044.47+222646.9 	& 0.024 &	2018-01-17 & 	KP	&	2022-03-31 & 	KP	&	NGC 3808	&		\\ 
SDSS J114222.33+084611.5 	& 0.022 &	2018-05-15 & 	JKT	&	2022-04-05 & 	KP	&	IC 720 	&		\\
SDSS J115805.22+275244.0 	& 0.011 &	2021-03-12 & 	JKT	&	2021-02-10 & 	KP	&	NGC 4004	&		\\ 
SDSS J120445.19+311132.9	& 0.025 &	2021-03-12 &	JKT	&	2021-03-10 &	KP	&		&		\\ 
SDSS J120607.51+325340.9 	& 0.027 &	2020-01-14 & 	JKT	&	2020-02-24 &	KP	&	IC 2993	&		\\ 
SDSS J121155.07+403918.2	& 0.022 &	2021-03-12 & 	JKT	&	2022-02-03 &	KP	&		&		\\ 
SDSS J121535.84+281039.4 	& 0.022 &	2018-03-13 &	JKT	&	2022-04-05 & 	KP	&	NGC 4211	&		\\
SDSS J121917.19+120058.2 	& 0.026 &	2021-04-05 & 	CT	&	2022-02-28 & 	KP	&	IC 3147	&		\\
SDSS J122257.72+103253.9 	& 0.026 &	2020-03-16 &	CT	&	2021-02-10 &	KP	&	NGC 4320	&		\\
SDSS J122814.15+442711.7 	& 0.023 &	2021-03-12 & 	JKT	&	2021-03-10 & 	KP	&	UGC 7593	&		\\ 
 SDSS J123108.37+003649.3 	& 0.023 &	2021-04-05 & 	CT	&	2022-03-31 & 	KP	&	NGC 4493	&		\\ 
SDSS J123634.26+111419.9 	& 0.008 &	2022-05-22 & 	KP	&	2022-03-27 & 	KP	&	NGC 4567/8	&		\\ 
SDSS J124109.46+293219.1	& 0.024 &	2021-03-12 & 	JKT	&	2022-01-31 & 	KP	&		&		\\ 
SDSS J124549.51+404904.2	& 0.024 &	2021-12-08 & 	KP	&	2021-12-08 & 	KP	&		&		\\ 
SDSS J124610.14+304354.9 	& 0.022 &	2021-03-17 & 	JKT	&	2021-02-16 &	KP	&	NGC 4676	&		\\
SDSS J125649.89+481800.4 	& 0.028 &	2019-06-28 & 	KP	&	2022-04-04 &	KP	&	NGC 4837	&	UGC 8068	\\ 
SDSS J125731.95+282836.9 	& 0.023 &	2019-06-19 & 	KP	&	2022-04-04 &	KP	&	NGC 4841	&		\\ 
SDSS J125818.23+290743.6 	& 0.026 &	2018-07-05 &  	JKT	&	2022-04-04 &	KP	&	IC 3935	&		\\
SDSS J125821.70+280855.5 	& 0.026 &	2022-04-05 & 	KP	&	2022-02-11 & 	JKT	&	NGC 4851	&		\\
SDSS J125940.10+275117.7	& 0.013 &	2022-05-01 & 	KP	&	2022-02-28 & 	KP	&		&		\\ 
SDSS J130004.24+280918.3	& 0.022 &	2022-05-02 & 	JKT	&	2022-02-03 &	KP	&		&		\\ 
SDSS J130125.26+291849.5 	& 0.023 &	2019-06-28 &  	KP	&	2018-05-15 & 	KP	&	NGC 4922	&		\\
\hline
\end{tabular}
\end{minipage}
\label{tbl-candidates}
\end{table*}

\setcounter{table}{1}
\begin{table*}
\contcaption{Sample galaxies and log of imaging observations}
 \centering
 \begin{minipage}{140mm}
   \caption{}
	\begin{tabular}{lcccccll} 
		\hline
SDSS name	&	$z$ & V date & V site	&	F510 date	& F510 site	&	Other ID	&	Notes	\\
  \hline
SDSS J130147.06+290435.9	& 0.023 & 	2022-05-02 & 	JKT	&	2022-03-27 &  	KP	&		&		\\  
SDSS J130702.74+133814.4 	& 0.027 & 	2021-03-12 & 	CT	&	2022-01-31 &	KP	&	KPG 365A	&		\\
SDSS J132035.40+340821.7 	& 0.023 & 	2018-06-14 & 	KP	&	2021-02-21 &	KP	&	IC 883	&		\\
SDSS J132652.11-023436.6 	& 0.012 & 	2021-04-15 & 	CT	&	2022-03-31 & 	KP	&		&		\\
SDSS J133525.40+332641.6 	& 0.024 & 	2020-07-24 & 	JKT	&	2022-04-04 &	KP	&		&		\\ 
SDSS J133817.77+481641.0 	& 0.028 & 	2021-03-17 & 	JKT	&	2021-03-17 & 	JKT	&	Mkn 266	& \cite{Keel2012} cloud	\\
SDSS J134139.62+554014.3 	& 0.025 & 	2015-03-27   &.  KP	&	2015-03-27    & KP.   &	NGC 5278	&	\cite{crossion} cloud	\\ 
SDSS J135837.97+372528.2 	& 0.012 & 	2021-03-17 & 	JKT	&	2020-02-21 & 	KP	&	NGC 5394/5	&		\\
SDSS J140037.96-025422.7 	& 0.025 &  2021-03-10 & 	KP	&	2021-03-10 &	KP	&	IC 968	&		\\ 
SDSS J140141.38+334936.5 	& 0.026 & 	2020-07-24 &  	JKT	&	2021-05-07 & 	KP	&	NGC 5421	&		\\ 
SDSS J140821.09+354221.3	& 0.028 & 	2022-04-05 &	KP	&	2022-02-03 & 	KP	&		&		\\
SDSS J140834.22+091147.7	& 0.023 & 	2021-04-15 & 	CT	&	2022-03-31 &	KP	&		&		\\
SDSS J141338.69+073935.8 	& 0.024 & 	2021-03-17 &	JKT	&	2021-03-17 & 	JKT	&	NGC 5514	&	TELPERION cloud	\\ 
SDSS J141342.00+081316.3 	& 0.025 & 	2021-04-15 & 	CT	&	2022-02-03 & 	KP	&	UGC 9103	&		\\ 
SDSS J141422.99+025831.5	& 0.026 & 	2020-03-16 & 	CT	&	2022-02-28 & 	KP	&		&		\\ 
SDSS J141631.80+393520.5	& 0.026 & 	2020-07-23 & 	JKT	&	2022-02-03 & 	KP	&	NGC 5541	&	DR7, not in DR12	\\ 
SDSS J141702.52+363417.7 	& 0.010 & 	2020-07-24 & 	JKT	&	2022-03-27 & 	KP	&	NGC 5544/5	&		\\ 
SDSS J141814.94+342659.8	& 0.028 &	2020-07-23 &	JKT	&	2022-03-03 & 	KP	&		&		\\ 
SDSS J142409.12+121531.1	& 0.027 & 	2021-04-15 & 	CT	&	2022-20-28 & 	KP	&		&		\\ 
SDSS J144042.83+032755.5 	& 0.027 & 	2019-06-03 & 	CT	&	2022-03-27 &	KP	&	NGC 5718	&		\\
SDSS J144514.14+384344.1 	& 0.015 & 	2018-03-13 & 	JKT	&	2022-04-04 & 	KP	&	NGC 5752	&		\\
SDSS J150448.33+410016.1 	& 0.028 & 	2020-07-23 &  	JKT	&	2022-02-11 &  	JKT	&		&		\\ 
SDSS J151100.89+233633.3 	& 0.017 & 	2020-07-24 & 	JKT	&	2022-03-27 &  	KP	&	MCG +04-36-024	&		\\
SDSS J152607.95+414033.8 	& 0.009 & 	2020-07-21 & 	JKT	&	2019-07-03 &  	KP	&	NGC 5929/30	&		\\ 
SDSS J153926.06+245636.9 	& 0.023 &	2020-07-23 & 	JKT	&	2019-05-04 & 	KP	&	Mkn 860A	&		\\ 
SDSS J155912.54+204548.0 	& 0.015 & 	2020-07-21 & 	JKT	&	2022-04-05 & 	KP	&	NGC 6027	&		\\
SDSS J162946.50+405229.5	& 0.028 & 	2017-09-28 & 	JKT	&	2019-07-03 & 	KP	&		&		\\ 
SDSS J215507.85-004613.0	& 0.027 & 	2021-10-06 & 	KP	&	2021-10-06 & 	KP	&		&		\\ 
SDSS J235410.08+002258.2 	& 0.026 & 	2021-10-06 & 	KP	&	2021-10-06 & 	KP	&	NGC 7783	&		\\
		\hline
	\end{tabular}
\end{minipage}
\end{table*}

\section{Results: EELRS in a large interacting and merging sample}

\begin{figure*}
	\includegraphics[width=140mm,angle=90]{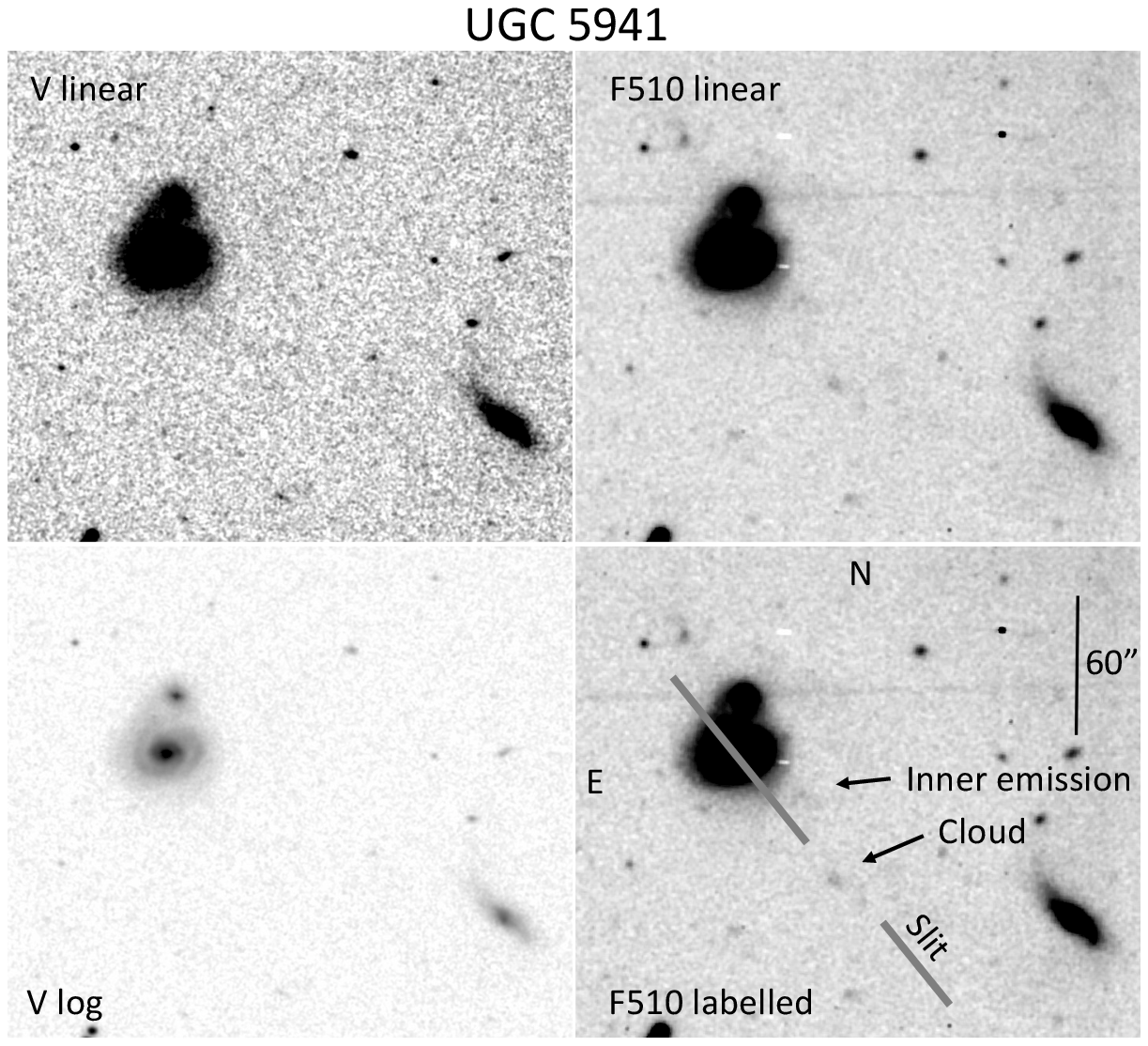}
    \caption{{Narrow- and broad-band} images showing the [O III] cloud detected near UGC 5941 (the galaxy pair at upper left). The area shown in 
    each panel spans
    208\arcsec N-S by 242\arcsec E-W, with north at the top as shown. Each image has been smoothed by a Gaussian with FWHM=0.65\arcsec 
    for display. The upper panels compare $V$ (predominantly continuum) and F510
    images with comparable linear stretch, showing the [O III] cloud to the SW and fainter structure just outside the bright disc. At lower left, 
    the $V$ image is shown with a log scale to reveal the spiral pattern of UGC 5941, while the F510 image is repeated at lower right with
    annotation including the location of the spectrograph slit.}
    \label{fig:UGC5941montage}
\end{figure*}

\begin{figure*}
	\includegraphics[width=85mm,angle=90]{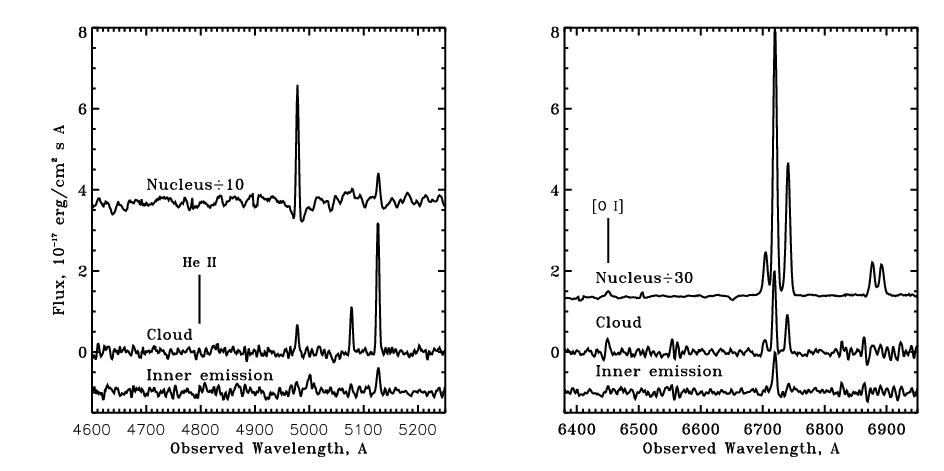}
    \caption{Diagnostic sections of the BTA long-slit spectrum of UGC 5941 and its associated emission cloud. The nucleus, summed over a 1 \arcsec 
    $\times 2$\arcsec region, is scaled down as indicated. The distant cloud spectrum is summed over a region 1\arcsec $\times 7.2$\arcsec, and the inner emission region is summed over $1 \times 7.6$\arcsec. Wavelengths of two weak or undetected lines which are important in classification are marked - He II $\lambda 4686$ in the blue
    segment and [O I] $\lambda 6300$ in the red segment. The [S II] doublet observed near 6870 \AA\  falls in a wavelength region so 
    contaminated by night-sky residuals that we can derive only upper limits for the cloud and inner emission region.}
    \label{fig:UGC5941specfig}
\end{figure*}

This survey recovered previously-known distant EELRs in three systems, and one probable new one. The three are Mkn 266 (with the highest-surface-brightness EELR from the Galaxy Zoo sample, \citealt{Keel2012}), the emission filaments beyond the stellar discs of NGC 5278/9 \citep{crossion}, and
the very distant cloud in the tidal tails of NGC 5514, found in the TELPERION AGN-host sample \citep{TELPERION}. 
The new [O III] structure was found near UGC 5941 (see Fig. \ref{fig:UGC5941montage}), at projected separation 70\arcsec or 35 kpc
from the nucleus. The BTA spectrum in Fig. \ref{fig:UGC5941specfig} shows ratios of the strong emission lines that place it in the AGN section of 
common classification diagrams, although its signal-to-noise ratio is not sufficient to detect He II $\lambda 4686$.
The upper limit to the ratio He II $\lambda 4686$/H$\beta < 0.25$
is high enough to be consistent with the values seen on other EELRs (\citealt{Keel2012}, \citealt{Knese}, \citealt{TELPERION}). 
This line ratio  has proven important in distinguishing the
ionization sources of gas at low metallicity, where the AGN and star-formation loci can overlap in the most common 
BPT diagram (\citealt{BPT}, \citealt{Kewley2001}, \citealt{Kauffmann}), as discussed by \cite{Groves2006}. It lies close in the BPT diagrams to previously-identified EELRs 
with He II detections.

Properties measured from the spectrum of the UGC 5941 cloud are listed in Table \ref{tbl-lineratios}, as are measures of
weaker and more diffuse emission just outside the galaxy disc along the slit. The redshifts are slightly smaller than that
we measure for the galaxy nucleus, although the radial-velocity differences are barely significant (listed under $v_{\rm offset}$
in the table,
in the sense object {\it minus} nucleus).

\setcounter{table}{1}
\begin{table}
	\centering
	\caption{Properties of UGC 5941 clouds}
	\label{tbl-lineratios}
	\begin{tabular}{lcc} 
		\hline
Property & Cloud & Inner Emission \\
\hline
~ $\alpha_{2000}$ & 10:50:17.69 & 10:50:20.00 \\
~ $\delta_{2000}$ &  +41:26:56 & +41:27:36 \\
~ $z$ & $0.02384 \pm 0.00005$  & $0.02363\pm0.00035$  \\
~ $v_{\rm offset}$, km s$^{-1}$ & $-34 \pm 17$ &  $-97 \pm105$ \\
~ [O III] $\lambda 5007/$H$\beta$ & $4.6 \pm 0.5$  & $>3,6$  \\
~ [N II] $\lambda 6583/$H$\alpha$ & $0.46 \pm 0.09$  & $0.19 \pm 0.12$ \\
~ [O I] $\lambda 6300/$H$\alpha$ & $0.18 \pm 0.05$ & $ < 0.18$ \\
~ H$\alpha$/H$\beta$ & $4.0 \pm 1.0$& $>5$  \\
~ [S II] $\lambda \lambda 6717,6731$/H$\alpha$ & $ <0.12$ & $<0.11$ \\
		\hline
	\end{tabular}
\end{table}

\subsection{Emission-line ratios and classifications}

\begin{figure*}
	\includegraphics[width=120mm,angle=90]{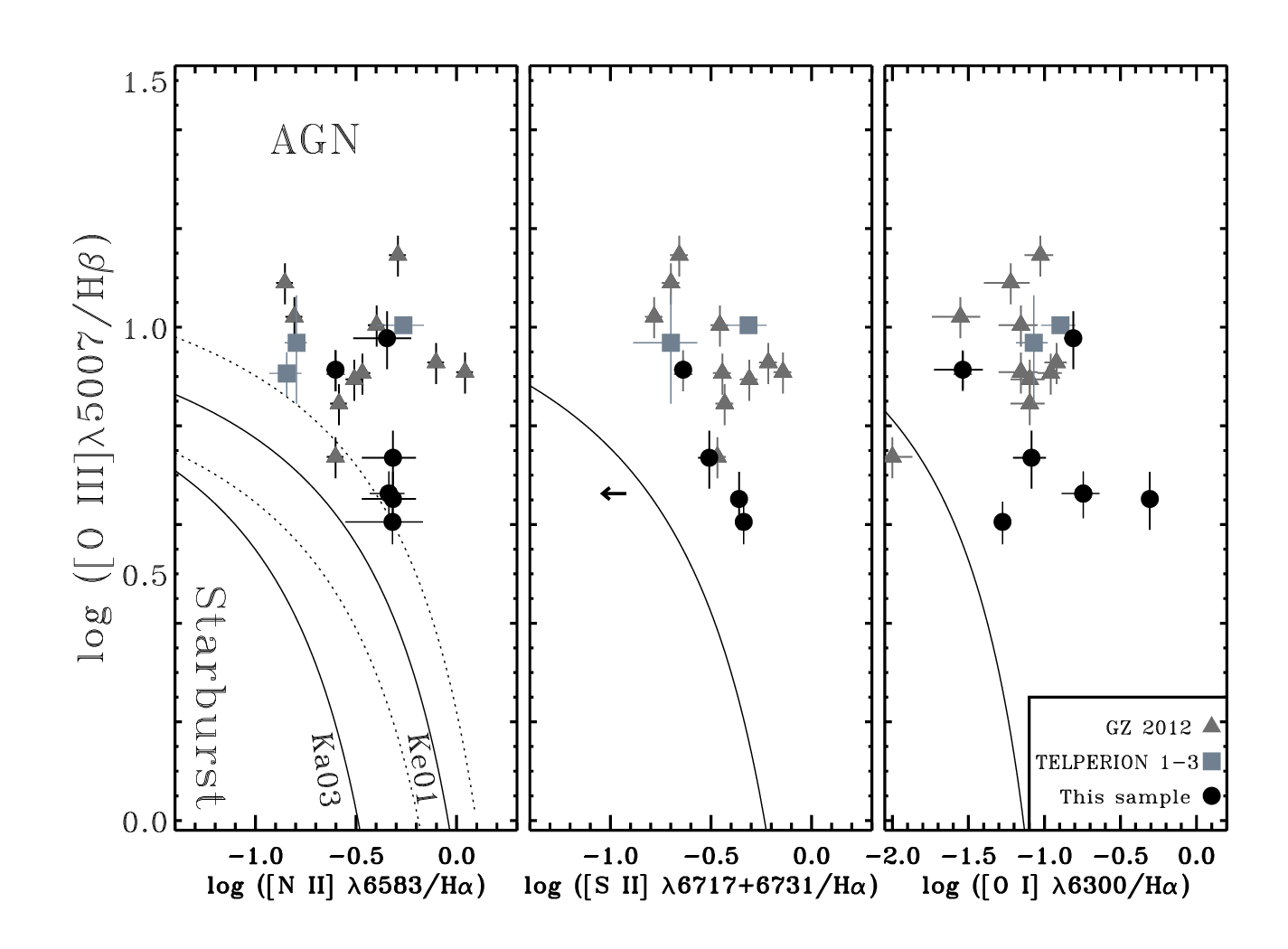}
    \caption{BPT emission-line classification diagrams for the EELRs in the Galaxy Zoo sample and the varioos phases of the TELPERION survey. Triangles indicate the Galaxy Zoo objects with red spectral coverage from \citealt{Keel2012}, squares mark objects from the earlier TELPERION phases and circles indicate objects from
 the sample observed for this work. The maximum-starburst line is adopted from \citealt{Kauffmann} (Ka03), and also shown according to \citealt{Kewley2001} (Ke01) for the leftmost diagram, 
with their recommended uncertainty of $\pm 0.1$ dex. Separate points are shown for three of the large filaments in the NGC 5278/9 system. We have taken values for NGC 7252 from \citealt{Schweizer} and for Mkn 266 from \citealt{Keel2012}. The right end of the arrow
in the [S II] diagram marks the upper limit for UGC 5941. Error bars for the Galaxy Zoo sample are representative, from objects with multiple observations.}
    \label{fig-BPTplots}
\end{figure*}

Common tools for understanding the energy input to gaseous nebulae, especially for galactic nuclei, are the
BPT diagrams \citep{BPT} comparing ratios of strong emission lines, as revised by later empirical and modeling studies (\citealt{VeilleuxDEO}, \citealt{Kewley2001}, \citealt{Kauffmann}).
The classic BPT classification diagrams are shown in Fig. \ref{fig-BPTplots} for the various phases of the TELPERION survey,
highlighting the objects appearing in this work, and for the Galaxy Zoo EELR clouds, including Hanny's Voorwerp near IC 2497 and sharing Mkn 266 in common 
with the sample observed for this paper. 
Only for UGC 5941 is there not a detection of He II at intensity $> 0.1$H$\beta$ to further support the identification of
AGN ionization; its cloud lies in the AGN region, above the maximum-starburst lines shown. The Galaxy Zoo objects as a group
have higher excitation (larger values of [O III]/H$\beta$) than the TELPERION clouds, attributable to higher ionization parameter and reflecting the deeper
surface-brightness threshold of the TELPERION narrowband survey compared to broad-band selection from SDSS images.

\subsection{Incidence of distant EELRs}

A straightforward way to estimate the incidence of distant EELRs is to consider the fractions of galaxies and known AGN hosts with such objects
in the various phases of the TELPERION survey. These include a sample of Seyfert galaxies with H I detections (H I AGN in Table \ref{tbl-tallies}), so that it would 
be known where
any ionized structures lie with respect to extended neutral gas \citep{Knese}; an extended version of the Toomre merging sequence (Toomre seq), a sample
of AGN hosts selected on luminosity \citep{TELPERION}; and the present work selecting a larger sample of strongly interacting and merging galaxies (Mergers). 
For completeness, we list the number of galaxies in each sample (Targeted) and the number of galaxies examined for EELRs including any additional
comparably luminous galaxies at matching redshifts (All); no EELRs appeared near these additional galaxies.

Simple tallies of the distant EELRs we have detected in the various TELPERION samples are
given in Table \ref{tbl-tallies}, where values for combined samples {account for galaxies in common between samples to avoid double-counting. These occur because the sample selections for various phases of the TELPERION survey used different kinds of criterion, and some galaxies satisfy more than one so were included in multiple samples.
This interacting and merging sample selected in this work overlaps with the Toomre-sequence galaxies from \cite{TELPERION} in the systems NGC 2623, NGC 4676, and IC 883,
and has NGC 5394/5, NGC 5514, and UGC 7064 in common with the AGN-host sample.
In assessing the detection statistics, we note further overlaps between the TELPERION samples as follows: The H I AGN list \citep{Knese}
and  AGN-host list from \cite{TELPERION} share the four systems UGC 3995, NGC 7469, NGC 7591, and NGC 7682.

We consider the numbers of galaxies, systems, and Seyfert galaxies.
Notably, for the Toomre sample, the EELR-to-Seyfert galaxy ratio is undefined due to the unique case of NGC 7252, which has an EELR without
directly-observed nuclear activity; \citep{Schweizer}. Various catalog sources differ in their thresholds for listing 
Seyfert nuclei as distinct from some LINERs, especially among the heterogenous sources used in compiling the VCV AGN catalog \citep{VCV}
used in constructing the AGN-host sample. Our values include the UGC 5941 cloud as AGN-ionized, based on its
location in Fig. \ref{fig-BPTplots}.
Uncertainties in the occurrence rates for distant EELRs
(EELRs/galaxy and EELRs/Seyfert) use the binomial expression\footnote{where the 1-$\sigma$ uncertainty is $(\frac{p (1-p)}  {N})^{1/2}$ for sample size $N$ and fractional occurrence $p$.}, since the number of galaxies observed was set independently
of the number of cloud detections.

LINERs counted here all have strong enough H$\alpha$ emission (equivalent width $>3$ \AA\ ) to be in the AGN family, except for shocked gas in ongoing mergers as in NGC 5514 \citep{Lipari} and NGC 6240, although even in this very dusty ultraluminous infrared galaxy (ULIRG) system there is evidence for AGN photoionization
in some directions \citep{Medling}.

\begin{table*}
	\centering
	\caption{Galaxies and Distant EELRs in TELPERION Samples}
	\label{tbl-tallies}
	\begin{tabular}{lcccccccl} 
		\hline
				Sample & \multicolumn{2}{c}{Galaxies:} & \multicolumn{2}{c}{Nuclei:} &  EELRs & EELR/galaxy & EELR/Seyfert & Reference \\

 & Targeted & All & Seyfert  & LINER & \\
\hline
H I AGN        &     26   &  34  &     21 &   5     &  1  & $0.029\pm 0.028 $& $0.048\pm 0.041$  &   \cite{Knese} \\
Toomre seq    &      17   &  21    &    0  &  9     &  1   & $0.047\pm 0.046$ & ---  & \cite{TELPERION} \\
AGN hosts       &   120   & 185    &   87   & 30   &    2  &   $0.011\pm 0.007$ & $0.023\pm 0.013$ &\cite{TELPERION} \\
Mergers         &    92   & 198     &   9  & 31    &   4   &  $0.020\pm  0.010$& $0.44\pm  0.05$ &  This work \\
Toomre seq+merger & 106 &   215     &   9 &  35     &  5 & $0.023\pm 0.010$ & $0.56\pm 0.17$ &  \\
H I AGN+AGN host    &   141 &   211   &   103 &  32    &   3 & $0.014\pm 0.008$ & $0.029\pm 0016$ &   \\
Interacting AGN &  34 & 34 & 34 & 40 & 5 & --- & $0.12\pm 0.06$ & \\    
All              &  241 &   415          & 111 &  71 &   7 & $0.017\pm 0.006$ & $0.064\pm 0.023$  & \\
		\hline
	\end{tabular}
\end{table*}


Whether counting per galaxy or per system, and whether we consider only interacting and merging galaxies or wider samples,
we find distant EELR clouds near 2-4\% of them. This contrasts with the Galaxy Zoo sample taken from a much larger set of galaxies
(roughly 16,000 potential AGN hosts) but with a surface-brightness limit in [O III] emission several times higher. That
survey found 19 clouds for a fraction of order 0.001. The redshift bounds for the samples in the TELPERION survey enforce
effective limits on galaxy luminosity; for the previous phase of the survey, AGN hosts were observed only if brighter than
absolute magnitude $-20$. We find occurrence fractions in TELPERION 20-40 times higher than the Galaxy Zoo sample, which may suggest that the number of EELRs climbs steeply toward lower surface brightness, so that even deeper imaging surveys might prove very fruitful. 

Our results continue the pattern of distant EELRs occurring preferentially in interacting and merging systems, as was seen in the Galaxy Zoo
objects. Among galaxies with Seyfert nuclei in the Toomre sequence sample and the merger sample from this paper, roughly half ($57\pm 17$\% ) show distant EELRs. We can extend the statistics slightly by including interacting AGN (Seyferts and LINERs)
from the AGN-host and H I AGN samples, taking the labels for tidal disturbance from Table 7 of \cite{TELPERION} and those listed
by \cite{Kuo} in their Group I, where H I features trace tidal interactions. Excluding duplications with the other TELPERION samples,
this addition comprises 25 Seyfert galaxies, 5 LINERs, and two EELRs near Seyfert galaxies. As shown in the ``Interacting AGN" entry in 
Table \ref{tbl-tallies},
this category includes all but one of the EELRs found in the whole TELPERION program. EELRs have been detected near $12\pm 6$\%
of this combined sample (and none of the noninteracting AGN observed).


\subsection{Galaxies and AGN with distant EELRs}

Combining the results of the four TELPERION phases, it is appropriate to compare the nuclei and clouds of each of the 
7 systems with distant EELRs. Fig. \ref{fig-o3montage} provides a visual comparison of the TELPERION EELRs, 
including UGC 5941, at the same scales in linear size and [O III] surface brightness. These systems are discussed starting with the
four found in this paper's merger sample.

{\bf Mkn 266 (NGC 5256)} is part of a merging system, whose [O III] clouds have the highest surface brightness among detections in the Galaxy Zoo survey
\citep{Keel2012}. With far-IR luminosity $10^{45}$ erg s$^{-1}$, it qualifies as a luminous infrared galaxy (LIRG). Both nuclei
show Sy 2 spectra in the optical range; this is one of the rare local examples of a dual AGN, with components separated by 5.6 kpc in projection
\citep{Mazzarella2012}. This, together with the ongoing merger and consequent deep blending of the nuclei in far-IR observations, makes it difficult to apportion the luminosity between the two galaxies to tell whether one nucleus dominates the photoionization.
Further, an ongoing starburst accounts for roughly half of the far-IR (and therefore bolometric) luminosity (\citealt{Wang1997}, \citealt{Mazzarella2012}). Mkn 266
is part of the merger sample reported in this paper, and also appeared in the Galaxy Zoo sample \citep{Keel2012} due to the very high surface brightness of its EELR filaments.

{\bf NGC 5278/9 (Arp 239, VV 19, Mkn 271)} form a strongly-interacting pair of spirals, which \cite{VV1977} called the ``Telephone Handset" system, noting a now-dated resemblance. Kinematic studies show the two distinct rotating discs which are prominent morphologically, but departures from circular coplanar motion are locally almost as large as the circular velocities 
(\citealt{Klimanov}, \citealt{Repetto}.) Simulations suggest that this combination, and the lack of bright tidal tails seen in the starlight, mark a galaxy pair on an initial  
close approach before the main bodies spiral inward
toward merger.
Both nuclei have similar optical spectra, near the Seyfert/LINER
boundary in BPT diagrams depending on aperture size. Each is accompanied by highly-ionized filaments extending
to projected distances 28-58\arcsec from the nuclei \citep{crossion}; the spectroscopic values we use here were obtained from new measurements on the data
obtained for that paper. Here again, the far-IR data blend both galaxies, and assessing how
much AGN radiation may be reradiated by dust is further complicated by the evident levels of star formation in the discs. We have explored the
IR characteristics of this system using the {\it Spitzer} data obtained in the Star Formation Reference Survey
(SFRS; \citealt{SFRS}). The SFRS project used the consistent continuum slope of a wide range of stellar populations across the mid-infrared bands, from 3.6-8$\mu$m
and therefore well longward of the generic 1.6$\mu$m peak in their spectral energy distributions,
to generate images which largely remove the direct light of stars at 8$\mu$m, by subtracting an appropriately scaled version of the 3.6$\mu$m image.  The resulting
image, their ``NS" combination, is especially powerful in isolated both warm dust in star-forming regions and emission from AGN. This technique applied to
NGC 5279 shows a central source at 8$\mu$m that is 4 times as bright as the core of NGC 5278, in contrast to their more equal
fluxes in narrow (optical) emission lines (where NGC 5279 is brighter in [O III] by a factor 1.5).

{\bf NGC 5514} is an advanced merger with two prominent tidal tails. The EELR, initially reported by \cite{TELPERION}, is near the tip of one tail,
at the extraordinary projected distance 75 kpc. Detailed analysis of that work's optical spectra in the main body, including the presence of
multiple superimposed velocity components, suggests that shock ionization is important here, so that scaling from the emission-line luminosity will overestimate 
the
AGN luminosity. Integral-field spectra suggest that off-nuclear starburst regions and associated
outflows dominate the shock emission \citep{Lipari}, rather than the large-scale shocks sometimes found in ongoing mergers. Unlike most of the
other LINER systems we deal with, in this case the line ratios must largely reflect shocks. In contrast, the very distant EELR is in a region
near the end of a stellar tail where large-scale cloud collisions are unlikely, shows small line widths and notably higher ionization levels
than gas near the galaxy center, all of which suggest photoionization by a (possibly faded) AGN. The direct emission signature of such an AGN
is lost in the complex environment of other processes near the galaxy nuclei, providing us only with an upper limit to AGN luminosity from available data.

{\bf UGC 5941} is an M51-type interacting pair, with a smaller companion projected 25\arcsec away near the tip of a tidal arm; the pair is accompanied by 
a similarly luminous edge-on spiral (SDSS J105008.59+412638.8) projected 162\arcsec from UGC 5941 at a similar redshift.

{\bf Mkn 1 (NGC 449)} shows an opposing pair of ionized clouds projected $\approx 13$ kpc from the nucleus found in the initial H I AGN phase of 
TELPERION \citep{Knese}. One of these clouds has
spectroscopic data of quality sufficient to show a high He II $\lambda 4686$/H$\beta$ ratio indicating AGN photoionization. The
host galaxy is in a pair with the spiral NGC 451 at projected separation 109\arcsec (equivalent to 35 kpc).
\cite{Kuo} find a common H I envelope including a tidal bridge and tails, suggesting that the
asymmetry and kinematic disturbance of the H I in the disc of Mkn 1 (also noted by \citealt{Omar} from GMRT data) have been caused 
by this interaction. The nucleus shows a typical Sy 2 optical spectrum.

{\bf NGC 235}, also designated NGC 235A in NED, is part of a tidally-distorted interacting triple including NGC 232 and the smaller galaxy NGC 235B, 
seen projected against the disc of NGC 235. The EELR \citep{TELPERION} lies between NGC 235 and NGC 232, whose nuclear
optical spectrum shows star formation without a significant AGN contribution. 

{\bf NGC 7252}, part of the original Toomre sequence, is a merger remnant in an advanced state with a single nucleus,
small-scale spiral pattern with star-forming regions, and extensive tidal tails. Its EELR was reported by \cite{Schweizer},
and recovered in the TELPERION merging-sequence sample \citep{TELPERION}. The ionized gas roughly follows one of
the tidal tails over a radial range 5-13 kpc from the nucleus. This system is notable for having no detected AGN, furnishing
a very strong case for essentially complete shutdown of accretion luminosity over $1-2 \times 10^4$ years.

Since many of the distant EELRs from the Galaxy Zoo sample are so bright as to suggest that the AGN has faded
over $10^4 - 10^5$ years, we consider the available data for the TELPERION objects to see whether some show evidence for fading.
(AGN brightening is more difficult to demonstrate, since it has the same energy-balance signature as lack of extended neutral gas
or greater-than-average AGN obscuration). For the luminous Galaxy Zoo AGN examined by \cite{Keel2012}, a simple picture 
could be used to bound the ionizing luminosity, in which all ionizing radiation leaving the nucleus either ionizes observable gas
(estimated using scaling relations from [O III] luminosity) or is absorbed and reradiated by dust, so the mid- and far-infrared
luminosities are proxies for the remaining fraction of ionizing radiation emitted from the AGN. As we find EELRs associated with
weaker (and sometimes unobservably faint) AGN, these assumptions become progressively less realistic. The contribution to grain heating from star
formation becomes relatively more important for lower AGN luminosity, and a larger fraction of the observed infrared fluxes for lower angular resolution. Likewise,
lower-luminosity AGN may have signatures in the optical spectra that cannot be reliably separated from central star formation using only the strong emission lines
incorporated in the BPT diagrams. The TELPERION
objects are clearly in this more complex regime.

This complexity may be seen from the data in Table \ref{tbl-fadestatus}, which includes information on the implied [O III]
luminosity of each nucleus, and its bolometric luminosity under the assumption that the entire far-infrared output comes from the nucleus,
as was done for the Galaxy Zoo sample. Here, L$_{\rm [O~ III]}$ is the luminosity of the nucleus in the [O III] $\lambda 5007$ line, while the bolometric luminosity L$_{\rm Bol}$ combines contributions traced by [OIII] and the far-infrared luminosity.  All these values of L$_{\rm ion,cloud}$  are in fact lower
limits, calculated for complete absorption of the incident ionizing continuum, so the more realistic case of incomplete absorption
increases the required luminosity of the AGN. The column labelled ``Ratio" is the ratio of this implied bolometric output to the 
ionizing luminosity L$_{\rm ion,cloud}$ needed to ionize the cloud at the observed H$\alpha$ surface brightness and (projected) distance from the nucleus, again estimated following \cite{Keel2012}; values smaller than unity would indicate that the AGN has faded over the relevant light-travel time. To the extent that surrounding
star-forming regions produce additional far-IR emission within the large beams of available FIR data, this ratio must
be an overestimate, sometimes by a very large factor. From these data, the AGN in NGC 5278/9, NGC 5514, and NGC 7252 show evidence of fading over timespans
0.9, 2.4, and 0.4$\times 10^5$ years, respectively. Mkn 1 is consistent with a constant luminosity obscured along our line of sight, while the situations in
Mkn 266 and NGC 235 remain ambiguous.

The case of the UGC 5941 EELR cloud illustrates some of the limits on this energy-balance approach for galaxies so distant that existing
mid- and far-infrared data do not separate emission from the nucleus and disk of the host galaxy. The AGN is faint enough to
be lost in the star-forming signatures of the galaxy both spectroscopically and in integrated infrared flux. Even if all the 
IR output of the galaxy is assigned to reprocessed AGN radiation, the cloud ionization still somewhat exceeds the level powered by an unobscured line of sight from the
cloud to the AGN, accounting for a fraction $<0.6$ of it from Table \ref{tbl-fadestatus}). UGC 5941 is therefore a candidate for modest fading over $\approx 10^5$ years.
The optical spectrum from the nucleus shows star formation, securely below the starburst dividing curves in all three BPT diagrams, with [N II]/H$\alpha$=0.48,
[O III]/H$\beta$=0.21. These ratios suggest that only a small fraction $<0.05$ of the H$\alpha$ or H$\beta$ emission at the nucleus comes from an unobscured AGN, based on a simple mixing model of two components typical of high-metallicity star-forming regions in galaxy nuclei and of Seyfert-galaxy line ratios.



We can start to address the limitations of IR sky survey data, improving luminosity estimates for low-luminosity AGN in star-forming host galaxies, for NGC 5278/9, using the 
{\it Spitzer} data from 3.6-24 $\mu$m obtained by the SFRS project \citep{SFRS}. This exercise sets tighter limits on the IR output from the
nuclei - at 8$\mu$m, the nucleus of NGC 5278 (NGC 5279) contributes no more than a fraction 0.034 (0.067) of the total system flux after correction for
stellar photospheric light, while at 24$\mu$m, the limit is 0.31 (0.23). If these values are representative of the system at longer wavelengths,
the ratios in Table \ref{tbl-fadestatus} should be divided by these fractions, which would push the NGC 5278/9 EELR clouds into the
regime requiring a fading AGN. Clearly this question will benefit from improved angular resolution throughout the infrared. An order-of-magnitude improvement in
discrimination of AGN from surrounding star formation at these redshifts, for example, would come from quite short exposures using MIRI on JWST.

Similarly, X-ray measurements can isolate AGN emission and provide strong constraints on the ionizing continuum, with appropriate caution
about the possibility of the source being strongly obscured at energies below $\approx 5 keV$ by surrounding material at high column densities.
For the EELR hosts in the TELPERION samples, the data do not yet exist to use this test at any greater sensitivity than spatially integrated
far-IR fluxes. The bright AGN in this group, NGC 235, Mkn 266, and Mkn 1, which are energetically capable of photoionizing the EELR according 
to the simple model described above, have X-ray detections that are compatible with this conclusion. NGC 7252, with no optical or IR
evidence of an AGN, has an X-ray luminosity and spectral parameters compatible with emission from its young stellar population and ISM
\citep{NGC7252XMM}. All the other TELPERION EELR hosts - NGC 5278/9, NGC 5514, and UGC 5941 - still lack even catalog X-ray
detections which are usefully deep for this purpose. Scaling from the bright detections suggests that point-source sensitivity near
$5 \times 10^{-14}$ erg cm$^{-2}$ s$^{-1}$ from 2-10 keV would show more conclusively whether, for example, all the clouds in the NGC 5278/9 system
show fading of the AGN, anchoring the SED of each AGN at energies where other sources are weaker and can be estimated from the
stellar population of the host galaxy.

\begin{table*}
	\centering
	\caption{Cloud-nucleus energy balance data}
	\label{tbl-fadestatus}
	\begin{tabular}{lccccccl} 
		\hline
Object        &	$z$ & Type &   L$_{\rm [O~ III]}$, erg s$^{-1}$ &    L$_{\rm Bol}$ &   L$_{\rm ion,cloud}$ & Ratio &  Projected distance, kpc \\
\hline
Mkn 1	 & 	0.016 & Sy 2 &   $3.0 \times 10^{41}$ &   $5.1 \times 10^{44}$ &    $ 6 \times 10^{42}$ &    <34  &    13 \\
NGC 7252 &	0.016  & ---    &     -   &    -               &         ---                           &                             &                           13 \\
NGC 235	 & 	0.022  & Sy 2 &  $ 1.3 \times 10^{41}$  & $2.8 \times 10^{44}$ & $ 2.4 \times 10^{44}$ &   <1.16   & 26 \\
NGC 5514 & 	0.024. &  Sy 2  & $1.1 \times 10^{39}$ &  $1.4 \times 10^{44}$ &   $ 3.9 \times 10^{44}$ &    <0.35 &   75 \\
Mkn 266	& 	0.028  & Sy2x2 & $6.2 \times 10^{40}$  & $1.1 \times 10^{45}$   & $ 3.9 \times 10^{44}$   & <2.8 &     25 \\
NGC 5278  inner & 0.025   &  LINER/Sy 2   & $6.8 \times 10^{39} $ &   $8.0 \times 10^{43}$ &   $3.2 \times 10^{43}$  &  <2.5 & 14 \\
NGC 5278     outer   &   0.025   &   LINER/Sy 2  &   $6.8 \times 10^{39} $ &  $8.0 \times 10^{43}$ &  $2.1 \times 10^{44}$ &  <0.38 & 29 \\
NGC 5279         &   0.025       &   LINER/Sy 2     &  $1.0 \times 10^{40}$ &  $8.1 \times 10^{43}$   & $2.9 \times 10^{44}$ &    <0.28 & 18 \\
UGC 5941       &  0.024        &  Star-forming      & $1.2 \times 10^{40}  $F$_{AGN}$ &   $ 2.7 \times 10^{44}$ & $4.7 \times 10^{44}$ & <0.6 & 35 \\
		\hline
	\end{tabular}
\end{table*}

\begin{figure*}
	\includegraphics[width=100mm,angle=0]{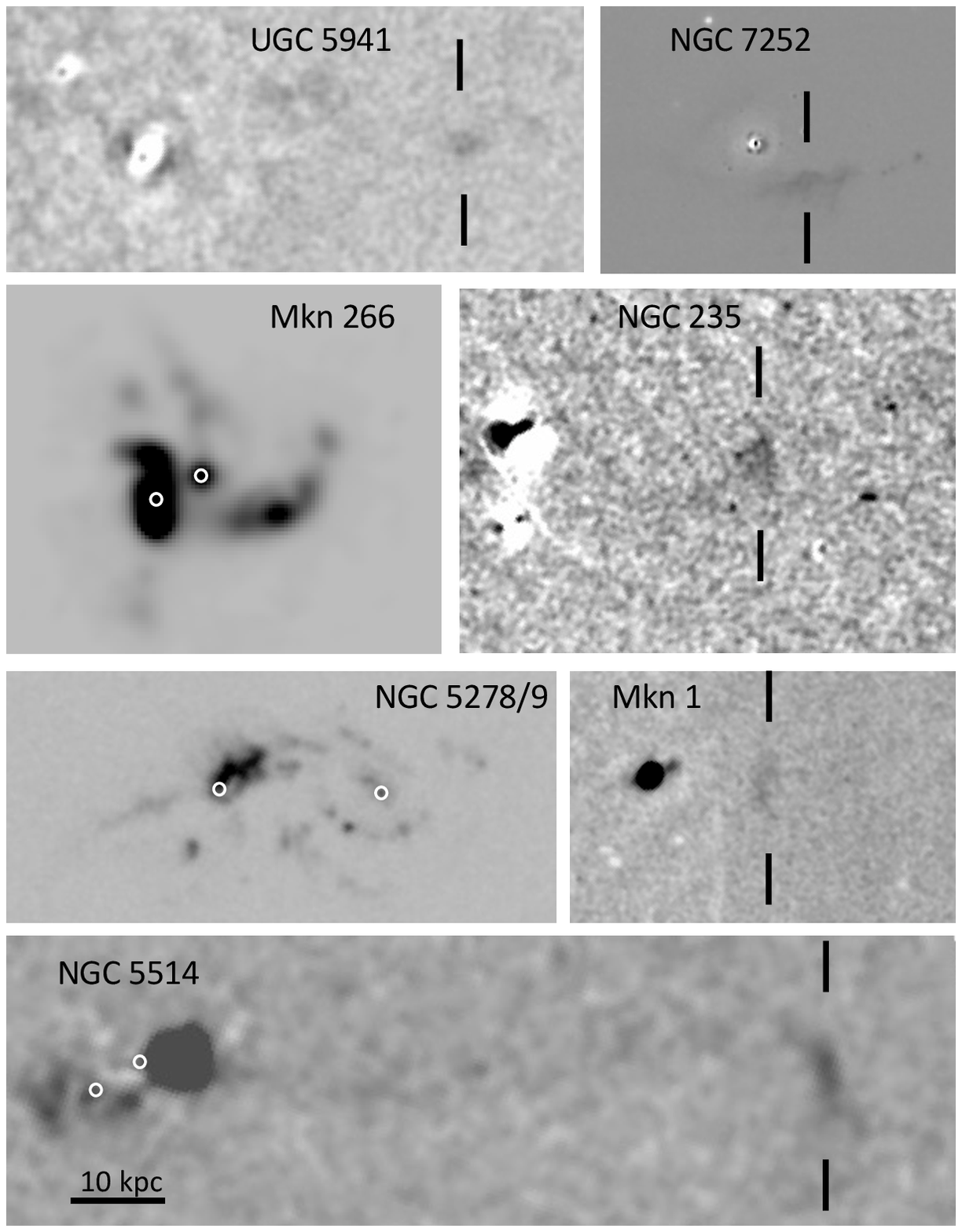}
    \caption{Comparison of the TELPERION EELRs, to the same scale in linear size and (approximately) the same [O III] surface-brightness
    scale. Mottling in the host-galaxy region of UGC 5941 is a result of imperfect matching of non-Gaussian PSFs. NGC 7252 is shown in data from the 2.5m DuPont telescope at Las Campanas, from \citealt{Schweizer}.
Data for Mkn 1 and NGC 235 are from SARA images. Mkn 1 shows artifacts due to imperfect removal of reflections from a bright star, but the ionization cones identified by \citealt {Stoklasova}
appear on either side of the nucleus. For Mkn 266, we show the intensity derived from fits through a Fabry-Perot data cube obtained with the 
SCORPIO-2 system at the BTA. The NGC 5514 and 5278/9 data used the tunable-filter MaNGaL system at the Caucasus Mountain Observatory
 2.5m telescope (section \ref{section:mangal}).
 Each image is rotated for convenient display. Small EELRs are indicated by tick marks above and below. For three systems with 
 multiple nuclei, their locations are shown by white circles.}
    \label{fig-o3montage}
\end{figure*}

\section{Conclusions}

We present the results of a narrow-band imaging search for distant EELRs (extending $>10$ kpc from the nucleus) around galaxies with and without AGN seen directly, to assess the luminosity history of the
AGN as well as contribute to our understanding of the pattern of emerging radiation from the cores. In a sample of 198 strongly-interacting and merging
galaxies within 92 systems, we have identified 3 previously-identified and one newly recognized distant EELR, with an AGN too weak to
separate from signatures of ongoing star formation but with the cloud lying in the AGN-ionized regions of BPT classification diagrams. Collecting these results with the three
previous TELPERION samples, we find detection rates of distant EELRs from 20-40 times higher than in the Galaxy Zoo
survey, which examined a much larger galaxy sample to [O III] surface brightness levels several times higher than reached with 
our current narrow-band images. Detection rates for distant EELRs around interacting Seyfert galaxies are particularly high, 
$12 \pm 6$\%, continuing the earlier preponderance of EELRs in interacting and merging systems. This must at least
in part reflect the presence of distant, often extraplanar H I in tidally distorted galaxies.

One of our motivations for the TELPERION survey was to assess the duration of bright periods in episodic AGN accretion. We have detected EELRs
around AGN sufficiently faint that the simple picture used earlier to ask whether AGN are more likely to be obscured or faded 
can no longer be applied with available data, because the integrated far-IR output is dominated by surrounding star formation
and X-ray catalogs are not yet deep enough. Analysis of this kind will need to await higher-resolution data in the mid and far-infrared or
deeper high-resolution X-ray observations.

\section*{Acknowledgements}

The narrowband filters were obtained thanks to a University of Alabama College of Arts and Sciences Dean's Leadership Board Faculty Fellowship.
Acquisition of new imagers for the SARA Observatory was supported by the National Science Foundation through
grant 0922981 to East Tennessee State University, and retrofitting of the Jacobus Kapteyn Telescope through grant 
1337566 to Texas A\&M University -- Commerce. We thank Steve Willner and Matt Ashby for copies of the processed SFRS {\it Spitzer} observations,
and Francois Schweizer for the Las Campanas [O III] image of NGC 7252.

Observations with the SAO RAS telescopes are supported by the Ministry of Science and Higher Education of the Russian Federation. The renovation of telescope equipment is currently provided within the national project ``Science and Universities". We obtained part of the observed data on the unique scientific facility ``Big Telescope Alt-azimuthal" as well as analysed the 2.5-m and 6-m telescopes data with the financial support of grant No075-15-2022-262 (13.MNPMU.21.0003) of the Ministry of Science and Higher Education of the Russian Federation.

The authors
are honored to be permitted to conduct astronomical research on Iolkam Du'ag (Kitt Peak), a mountain with particular significance to the Tohono O'odham Nation.
This research has made use of the NASA/IPAC Extragalactic Database (NED), which is operated by the Jet Propulsion Laboratory, 
Caltech, under contract with the National Aeronautics and Space Administration.
This research has made use of the VizieR catalogue access tool, CDS, Strasbourg, France (DOI : 10.26093/cds/vizier). The 
original description of the VizieR service was published in 2000, A\&AS 143, 23.

Funding for the Sloan Digital Sky Survey V has been provided by the Alfred P. Sloan Foundation, the Heising-Simons Foundation, the National Science Foundation, and the Participating Institutions. SDSS acknowledges support and resources from the Center for High-Performance Computing at the University of Utah. SDSS telescopes are located at Apache Point Observatory, funded by the Astrophysical Research Consortium and operated by New Mexico State University, and at Las Campanas Observatory, operated by the Carnegie Institution for Science. The SDSS web site is \url{www.sdss.org}.

SDSS is managed by the Astrophysical Research Consortium for the Participating Institutions of the SDSS Collaboration, including Caltech, The Carnegie Institution for Science, Chilean National Time Allocation Committee (CNTAC) ratified researchers, The Flatiron Institute, the Gotham Participation Group, Harvard University, Heidelberg University, The Johns Hopkins University, L’Ecole polytechnique f\'{e}d\'{e}rale de Lausanne (EPFL), Leibniz-Institut f\"{u}r Astrophysik Potsdam (AIP), Max-Planck-Institut f\"{u}r Astronomie (MPIA Heidelberg), Max-Planck-Institut f\"{u}r Extraterrestrische Physik (MPE), Nanjing University, National Astronomical Observatories of China (NAOC), New Mexico State University, The Ohio State University, Pennsylvania State University, Smithsonian Astrophysical Observatory, Space Telescope Science Institute (STScI), the Stellar Astrophysics Participation Group, Universidad Nacional Aut\'{o}noma de M\'{e}xico, University of Arizona, University of Colorado Boulder, University of Illinois at Urbana-Champaign, University of Toronto, University of Utah, University of Virginia, Yale University, and Yunnan University.

\section*{Data Availability}

The SARA images, and BTA long-slit spectrum of UGC 5941, are available at Zenodo.org under DOI 10.5281/zenodo.8374640. 
The {\it Spitzer} data may be retrieved from the Spitzer Heritage Archive at IRSA (https://irsa.ipac.caltech.edu/applications/Spitzer/SHA).




\bsp	
\label{lastpage}
\end{document}